# WEST ADVANCED WALL PROTECTION ACHIEVEMENTS TOWARD LONG PULSE OPERATION

R. MITTEAU[1,*], M.H. AUMEUNIER[1], L. DUBUS[1], J. GERARDIN[1], V. GORSE[1], E. GRELIER[1], V. MONCADA[1], S. VIVES[1], X. LITAUDON[1], M. JAKUBOWSKI[2], J. FELLINGER[2] and the WEST Team[3]

[1] CEA, IRFM, Saint-Paul-lez-Durance, France

[2] Max-Planck-Institut für Plasmaphysik, Greifswald, Germany

[3] the WEST team, see http://west.cea.fr/WESTteam

[*] Corresponding author, Email: raphael.mitteau@cea.fr

**Abstract**

Long pulse operation in magnetic fusion devices requires well controlled plasma power exhaust to the divertor & wall, and avoidance of wall hot spots that could evolve in wall damage. At WEST, 10 major plasma facing components are monitored using 10 series of temperature/power indicators, based on multiple diagnostic systems, among which the infrared viewing system is especially relevant. These indicators span from the most basic ones (temperatures, power and energy from deterministic models) to advanced processes using artificial intelligence acquired through machine learning. Some advanced processes do operate in real time, and feedback on power actuators through the plasma control system, providing active control toward remaining within the safe operational domain. Other advanced processes intervene as forensic tools post discharge to identify possible dangerous situations regarding the power loading to the wall, so that the discharge run plan is adjusted to avoid running into aggravating wall events. No critical wall power event happened during the campaigns C9 to C11 (2024-2025), totalling about 13h of plasma, that would have affected the campaign plan. While it cannot be demonstrated that the active & intelligent wall protection enabled the new plasma duration record of 1337 seconds, the wall protection system as a whole plausibly contributed to obtaining these record durations by preventing wall hot spots to become critical during the campaigns.

## 1. INTRODUCTION

Exploiting and extending WEST capability to long pulse operation represents a large part of WEST's experimental activity during the campaigns C9 to C11 (2024-2025), totalling about 13h of plasma [1]. High power Long Pulse Operation (LPO) in magnetic fusion plasma devices requires well-controlled plasma power exhaust to the In Vessel Components (IVCs), essentially divertor & main chamber wall. Any deviation to the carefully engineered power exhaust scheme results in wall hot spots and possible damage to the components. This effect is exacerbated by plasma duration, because LPO is enabled by employing actively cooled components, with a specific operation mode. The operation domain of cooled components is primarily limited by the surface heat flux (q, [W/m²]) and power (P, [W]), in addition to the standard & customary temperature limits commonly used for radiation cooled components. Flux and power quantities govern armour delamination and coolant critical heat flux in water-cooled components, while these flux & power quantities are much less of a concern for non-actively cooled IVCs. LPO requires a control over the engineering heat fluxes "q" & power "P" at the component surface. But no diagnostic produces real measures of engineering "q" or "P". Hence, having alternative, "good enough" estimates of the fluxes & power is essential. Moreover, actively cooled components have comparatively more severe damage modes than radiation cooled components, because of the possible breach to the cooling circuit. This exacerbates the need to remain strictly within the allowed operation domain, and at the same time renders the detection of abnormal wall thermal events essential for wall protection.

At WEST, 10 major IVCs (lower and upper divertor, lower baffle, inner guard limiters, five Lower Hybrid (LH) and ion cyclotron heating antennas, dump plates, Fig. 1) are monitored using 10 series of power / temperature indicators, largely based on the image streams from the InfraRed (IR) viewing system. They span from the most basic ones (heat fluxes, temperatures, power and energy from deterministic models) to advanced processes using neural networks and machine learning techniques. Some processes operate in Real Time (RT) and feedback on actuators through the Plasma Control System (PCS), providing active strategies towards remaining within the operational domain [2]. Other processes intervene post discharge as forensic tools to identify possible deteriorations of the wall and divertor under power loading. The quantitative power and temperature estimations from these processes are used for the preparation of the following discharge, possibly adjusting the discharge parameters to move away from wall security thresholds.

The infrared viewing system is a key contributor to IVCs protection. This is because of its unique combination of being a quantitative measure, being produced as a 2D + time data stream. The IR viewing systems provide the radiance of the hot thermal scene of the IVCs. WEST is equipped with 6 IR viewing systems which main objective is wall safety, while other IR viewing systems are used mostly for science and are not further mentioned here. The WEST infrared system is thoroughly described in [3].

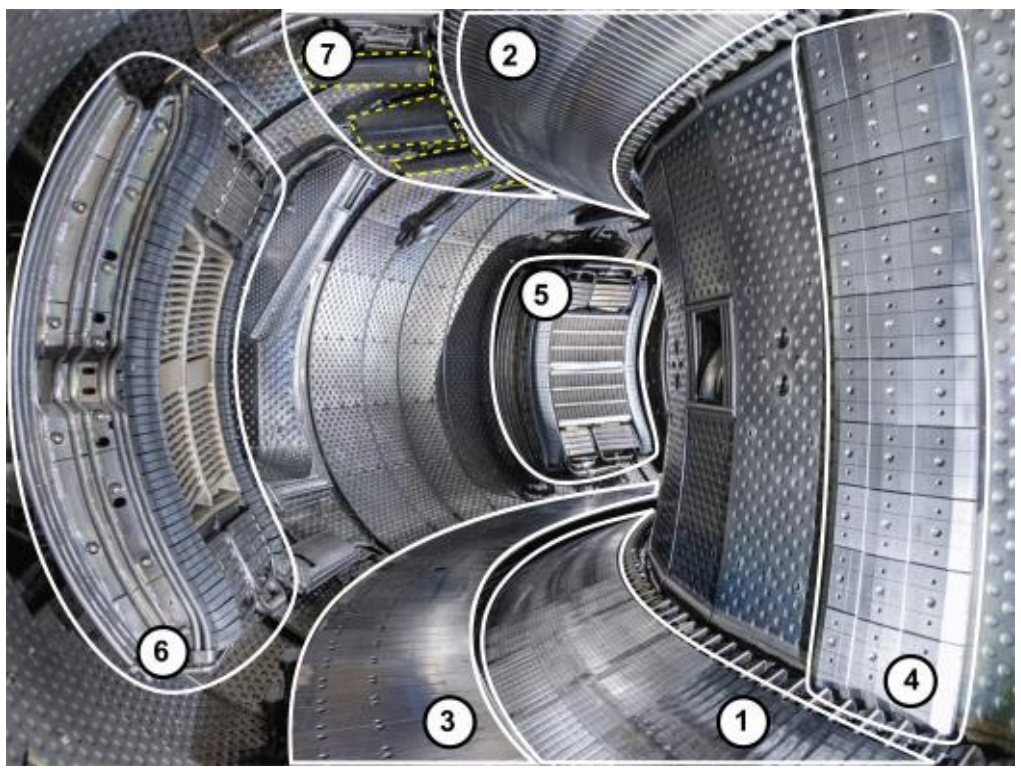

Fig. 1: WEST inner vessel (2024) with 7 of the 10 major internal components being monitored in infrared: 1) the lower divertor, 2) the upper divertor, 3) the lower divertor baffle, 4) the inner guard limiter, 5) the LH2 lower hybrid antenna, 6) the Q1 ICRH antenna, 7) the VDE and ripple dump plates.

While unrivalled as a wall health monitoring tool by the human wall monitoring experts (Plasma facing components Protection Officers, PPOs), the infrared images are complex to interpret. Obtaining the surface temperature from the scene radiance requires having sufficient knowledge of the thermal scene surface properties (emissivity ε, reflectivity r, possible presence of thermally resistive layers which increase the apparent surface temperature, compared to the temperature of a clean component under the same heat flux density). All these properties are variable with location and time, and photonic properties such as ε and r depend also on the light incidence angle. This complicates a lot any attempt at correcting the radiance toward obtaining the real unperturbed radiance. The conversion of the thermal scene to a surface temperature map is also affected by the predominantly diffusive or reflective nature of the wall. Possible hot emission sources outside of the Field of View (FoV) do affect the scene radiance because of reflections, and should normally be corrected too. Obtaining a good enough estimation of IVCs surface temperature gets tricky when ε becomes very low, which is the case in devices with metallic walls. As an example, a WEST divertor monobloc being observed at an apparent temperature of 454°C ("apparent" means ε taken to 1.0, i.e. blackbody temperature) is actually at 508°C for ε = 0.7 (high emissivity, representative of rough deposit), at 567°C for ε = 0.5 (average emissivity), at 675°C for ε = 0.3, (low emissivity), and finally 1026°C for ε = 0.1 (very low emissivity, although still within the actually tungsten blocks emissivity observations [4]). These temperature calculations with variable emissivity are made using the infrared analysis software ThermaVIP [5, 6], and involve the elaborated calibration laws of the WEST infrared viewing system [7]. From these numbers, the temperature error associated to an erroneous emissivity estimation can reach 500°C, about the same value as the temperature estimation itself, hinting at the very large error that can be associated to an improper emissivity assumption.

Beyond the emissivity challenge, converting the temperature time history into the equivalent surface heat flux time history requires temperature to flux inversion. This inversion is made uncertain by many uncertainties and complexities (non-linearities in material and heat transfer properties, complex geometries including armour castellations and segmentation, possible presence of thermally resistive layer at the surface of the IVCs, in addition to the uncertainties associated to ε and r already cited above). A comprehensive effort is underway at WEST, to untangle all these effects, though the use of inverse methods and synthetic diagnostic simulation, associated to Artificial Intelligence (AI) support tools [8]. This innovative method is due to be integrated in the WEST safety chain through a vessel thermal map, once mature enough. For the present purpose of machine safety, the estimation of the true component temperature is obtained by assuming a constant emissivity for each component, and neglecting the diffuse light reflexion. The emissivity is adjusted regularly, typically during yearly WEST maintenance shutdowns. The revised emissivity is obtained either by parameter estimation from cross diagnostic analysis or by in-vessel direct measures using a commercial equipment.

The basic wall safety system relies merely on monitoring the surface temperatures in prescribed Regions Of Interest (ROIs). They are compared to temperature limits, and, before reaching the limit, a controlled temperature band is set which feeds back on the injected power of the additional heating systems. These temperature traces are shown in Fig. 2 for the record discharge #61299, along with the temperature control bands. The lower end of the band is represented with the dashed grey line, the upper end in continuous black. Above the maximum prescribed temperature, a hard-wired connection triggers a soft plasma termination, playing the role of a defence-in-depth safety net.

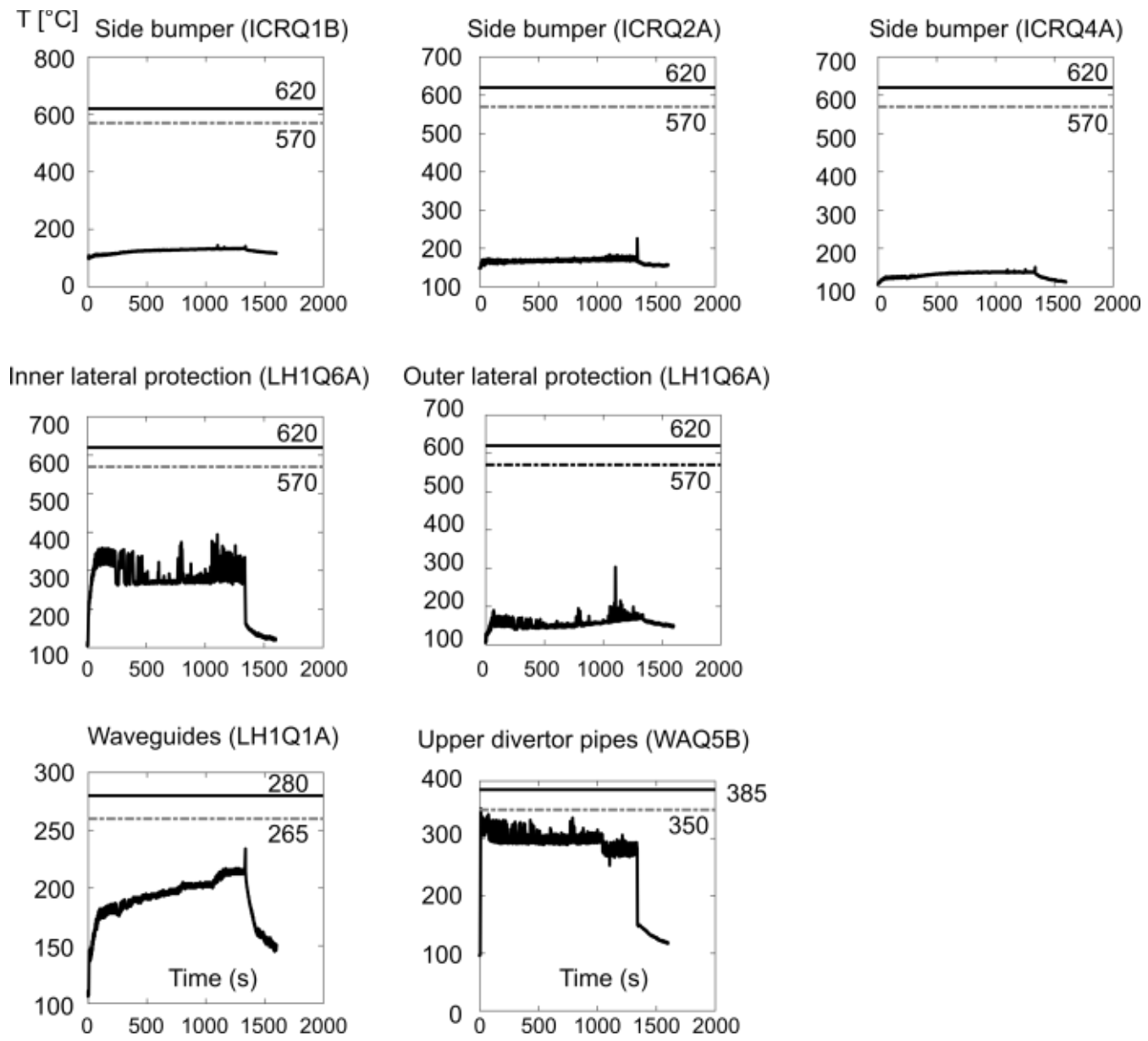

Fig. 2: Temperature plots (all ordinates in [°C]) of monitored regions of interest that define the operational domain of the LH antenna LH1 during the 1337 seconds discharge #61299.

In that example, all but one (the upper divertor pipes) stay within the allowable domain, while the LH1 waveguide heats up steadily during the whole pulse. The upper divertor pipes are heated by fast electrons from trapped ripple losses associated to the LH heating, a typical & controlled feature of these discharges. They enter the control band at the beginning of the discharge. Thanks to the feedback control and the water cooling, their temperature does not rise to the upper limit and finally decreases as the confinement increases, and the discharge goes on.

Such a security system based on ROI temperatures has many shortcomings:

— It is a temperature-based rather than a flux-based control, with the already above-mentioned limitations.
— Frequent and manual adjustment of ROI contours are needed, because of changes in the field of view (positional drift of the line of sight, or implementation of approved derogated regions).
— False alarms do happen, when an unexpected photonic reflexion or a hot chip of material passes through a ROI. Even though possibly short lived, the subsequent programmed reduction of injected power may trigger uncontrollable and finally deleterious MHD activity, shortening the discharge duration.

The practical conclusion is that certain safety actions should rather be triggered by specific geometrical shapes or patterns rather than a temperature. Better monitoring systems should: a) take into account the shape and/or the time sequence of the event, as visible on the IR frames; b) use other diagnostic measures relevant to IVCs, such as local temperature measures (thermocouples, fiber Bragg grating) [9], coolant calorimetry [10], and finally all machine diagnostic data. These various data streams should then be used and combined into intelligent indicators. Such combination is best done using engineering or physical models and/or Machine Learning (ML).

This paper focuses now on the 7 advanced wall protection processes that either rely on ML techniques, or cross-diagnostic data merging: the thermal event detector, the divertor strikeline descriptor, the LH heating system arc detector, the Runaway Electrons (REs) detector, the Unidentified Flying Objects (UFOs) detector, the post discharge automatic expert assessment of the wall thermal response, and the RT estimation of heat flux to the divertor from RT power balance. They are detailed hereafter.

2. ADVANCED WALL PROTECTION TECHNIQUES AND METHODS

### 2.1. Model based estimation of the incident heat flux on selected internal components

Having a solid estimate of the IVCs thermal flux density is key to IVC monitoring. The divertor engineering heat flux is obtained independently of the IR system in real time using a model-based estimator. The standard power distribution model is used, of which a comprehensive rationale is given in [11]:

$$q_{div}^{eng} = r_{in-out} \cdot r_{ripple} \cdot \frac{\boldsymbol{B_{OSP}} \cdot \boldsymbol{n_{Div}}}{\|\boldsymbol{B_{OSP}}\|} \cdot \frac{p_{Cond}}{4 \cdot \pi \cdot r_{OSP} \cdot \lambda_q^{OMP}} \cdot \frac{\|\boldsymbol{B_{OMP}}\|}{\boldsymbol{B_{OMP}} \cdot \boldsymbol{e_z}}$$

$\boldsymbol{B}$ is the total magnetic field. The subscripts OMP and OSP stand for Outer MidPlane and Outer Strike Point, respectively. $p_{Cond}$ is the conducto-convective power exiting the plasma ($p_{Cond} = p_{ohmic} + p_{add} - p_{rad}$), sometime referred to as being the scrape off layer power $P_{SOL}$. The plasma stored-energy term dW/dt is neglected in this power balance expression. This simplification is customarily done for plasma facing component power studies. It is either small compared to other terms of the balance, or short-lived when large. Being either small or short-lived, its contribution to the energy balance at the time and geometrical scales of the IVC tile is small. In the expression above, $r_{osp}$ is the major radius of the OSP, and $\lambda_q^{OMP}$ is the heat flux decay length of the scrape-off layer at the outer midplane. $\boldsymbol{n_{div}}$ and $\boldsymbol{e_z}$ are the unit vectors normal to the outer divertor target and to the horizontal plane, respectively. Compared to the expression given in [11], two WEST-specific correction factors are added: $r_{in-out}$ and $r_{ripple}$. $r_{in-out}$ is a scalar parameter expressing the in-out power deposition asymmetry ($r_{in-out}$ = 4/3), and $r_{ripple}$ is the peaking factor associated to the ripple of the toroidal magnetic field ($r_{ripple}$ = 2 for the WEST divertor). **Bold** quantities are vectors. $p_{ohmic}$, $\boldsymbol{B_{OSP}}$ and $\boldsymbol{B_{OMP}}$ are produced by the magnetic RT processing unit, $p_{add}$ by the additional heating systems processing units, and $p_{rad}$ by the bolometry processing unit. The heat flux decay length is prescribed at 7.5 mm, according to [12]. It has little variation over the customary WEST operation range, and the divertor heat flux for other decay lengths can readily be obtained by linear scaling. All other variable quantities are read by the wall monitoring system processing unit through a shared memory.

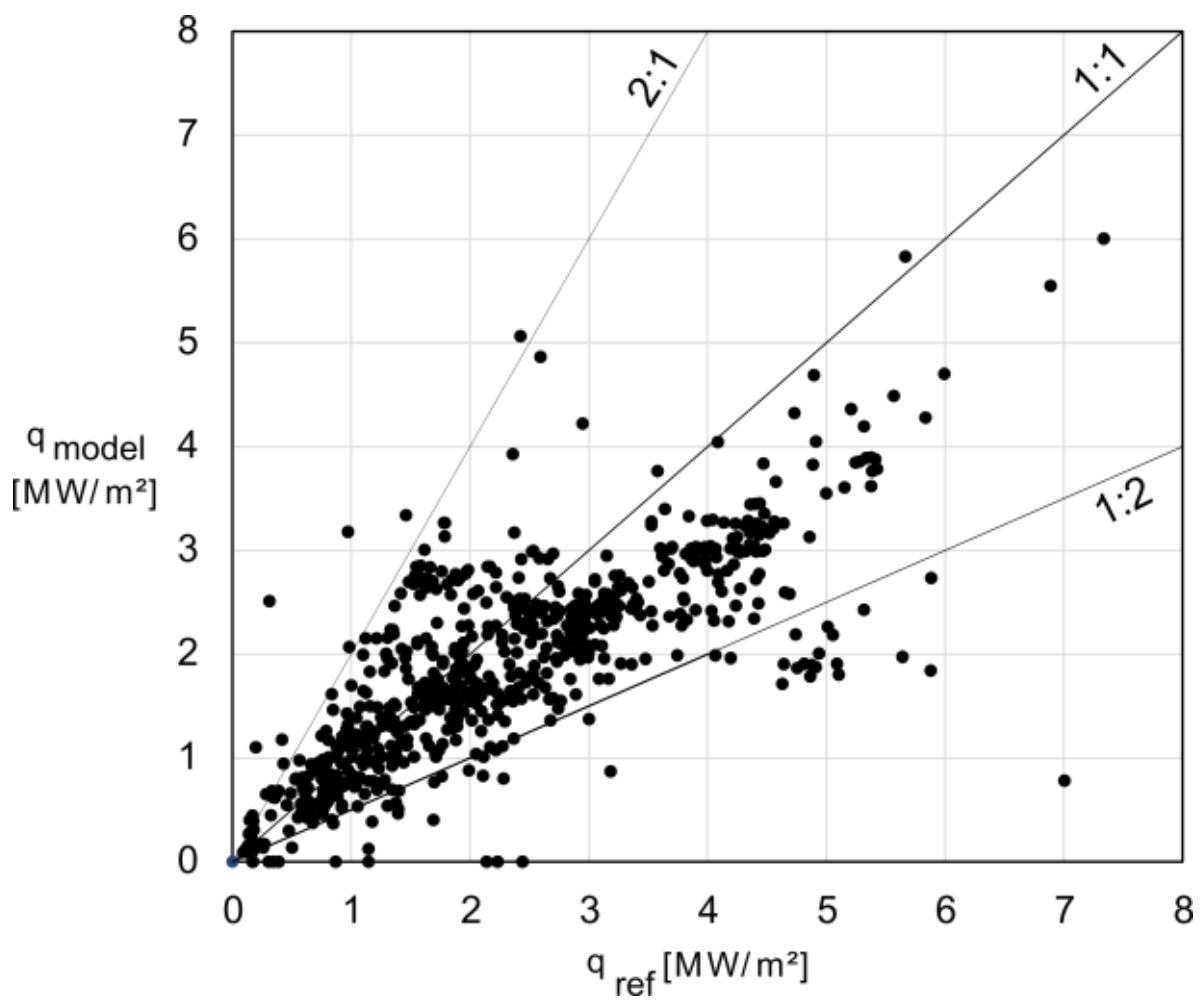

Fig. 3: Divertor engineering heat flux from the mode-based expression, as a function of inferred heat flux from embedded probes and inversed surface temperature measure from the IR viewing systems.

Note that the conducto-conductive power is adjusted by a proportional factor, as the WEST observed conducto-convective power is less than obtained from the power balance. This proportionality correction coefficient relates to the WEST thermal power balance, a recurring activity that is continuously followed and maintained up-to-date as reported in regular publications [10, 13, 14]. The correction factor was of 1.8 in 2018-2019, is currently of 1.6 and further reduces as the power balance improves, notably thanks to an improved calorimetry and power accounting of recessed wall components.This power distribution model is validated to experimental divertor heat flux estimations. This is done in relation to the activities of the WEST plasma wall interaction task force, of which significant resources are devoted to the estimation of the divertor heat flux. In this activity involving a cross disciplinary, multi-researcher' team, the divertor heat flux is validated from several diagnostics: IR thermography, embedded diagnostics (thermocouples, fiber brag gratings), Langmuir probes (both reciprocating probe and flush mounted probes) and finally calorimetry for a large variety of plasma modes and injected powers. Selected discharges are thoroughly documented and reported in [12, 15-18]. For given plasma conditions, the dispersion of the various divertor heat flux evaluations coming from the different evaluation paths already amounts to +/- 20% [12]. This activity results in a reference series of divertor heat flux measures, against which the model-based heat flux is validated (Fig. 3).

The resulting dispersion of the model-based heat flux is of a factor x0.5 to x2. This is acknowledged as being large, but deemed sufficient for using the whole process as a proof-of-principle. A follow-up activity is underway to improve the model-based divertor heat flux evaluation, by using a more accurate power distribution accounting for power being deposited to recessed components, and possibly by updating the power fall-off model from the simple exponential power decay to the Eich power decay model [18]. The divertor heat flux calculated from the model-based expression is written in a shared memory, and made available to other PCS processes. The RT calculated heat flux is shown at 2 MW/m² in Fig. 4, for discharge #61299. Its time variation follows closely the divertor power time trace. This engineering heat flux $q_{div}^{eng}$ is directly comparable to the divertor allowable of 10 MW/m² at WEST, and it stays well below that limit. This process operated flawlessly during C11 campaign.

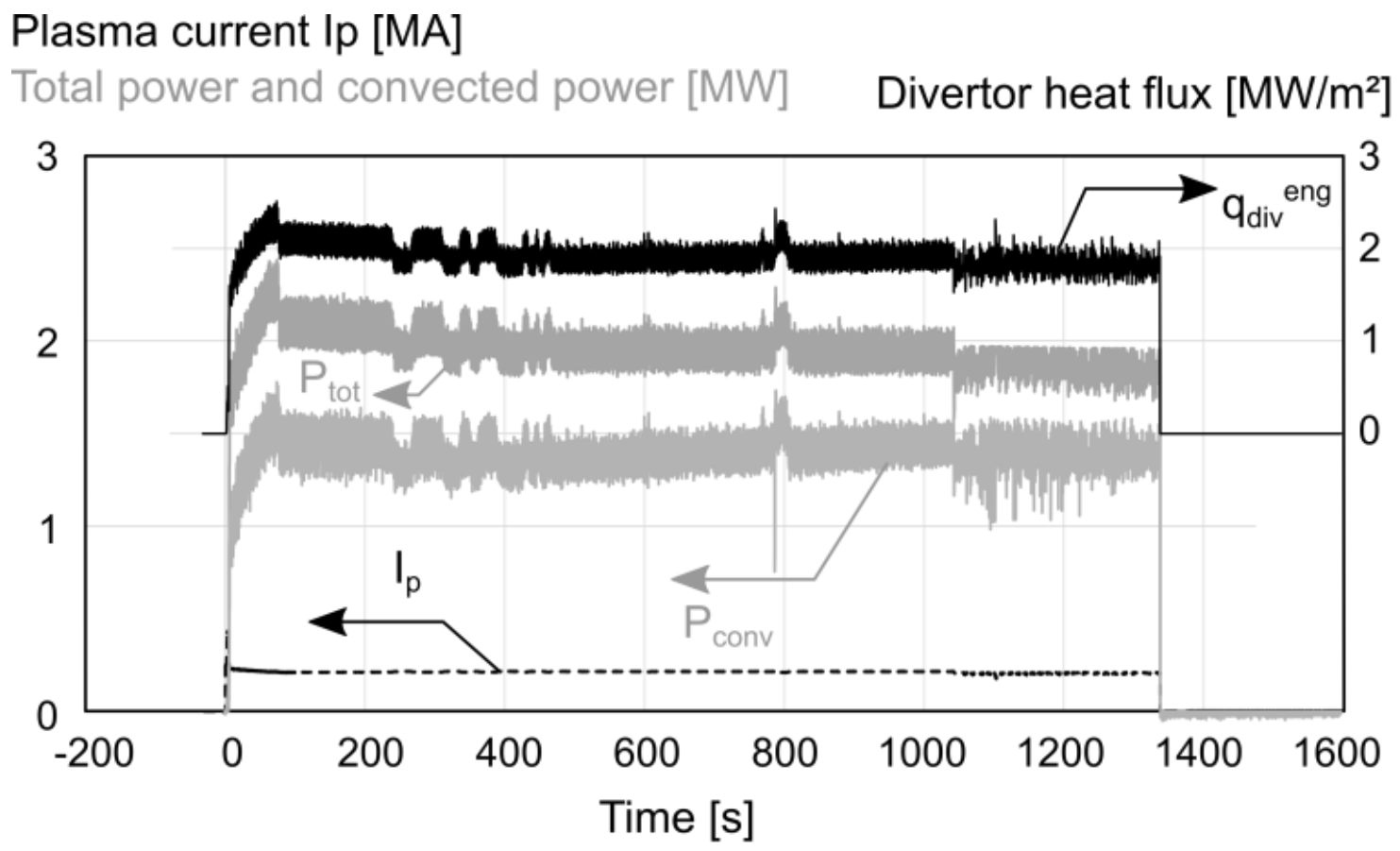

Fig. 4: Time traces of the RT calculated divertor heat flux and power from a RT power balance calculation, along with the measured plasma current, for the record discharge #61299.

Other power indicators are calculated from that basic power balance model. One noticeable indicator is the cumulated energy density to WEST vacuum vessel bellows, that heat up as a result of the radiation and of insufficient radiation screening from the IVCs. The bellow incident heat flux is derived from the separatrix radiated heat flux, using a geometrical form factor calculated from a radiation model. The maximum cumulated energy density on the bellow is obtained by combining the time-integrated bolometry heat flux combined with the radiation form factor. The cumulated energy on the vessel bellow stays well below the maximum allowable energy density of 2.1 MJ/m² during the long discharges.

### 2.2. Learning-based identification of thermal events

Better detectors are expected by introducing intelligence on the 2D image streams of the IR viewing diagnostic. The objective is to mimic the human operator's capability of interpreting the IR movies and pointing at potentially problematic thermal events, primarily according to their shape, location, and time evolution, rather than to the temperature of the event. This section addresses these processes based on machine learning: the thermal event, the strike line, the LH arc, the REs and the UFOs detectors.

The thermal event detector aims at detecting thermal events based on their shape or texture. The detector is built from a database of manually annotated events having occurred previously and available as a series of couples images + metadata. It is based on the concept of bounding boxes, a rectangle that surrounds the hot spot, being part of the metadata associated to the image. A time tracking module acting on the movie links the individual hot spots (associated to a single frame) to thermal events (which relates to a time sequence, characterised by starting and ending frame indices). Several annotation supporting tools have been built on purpose to generate these annotations. They facilitate greatly the generation of annotated databases, which are the cornerstone of AI developments. These developments of IR image-based advanced wall protection processes are done in an international framework, mainly within the EUFOfusion consortium. Thermal event formats (instances, categories, annotations) are developed jointly with IPP/W7-X [19, 20]. The ability to share data and knowledge is viewed as crucial, because cross-machine diversity is key to obtaining high-performance models. Most image data workflows used in both CEA/WEST and IPP/W7-X institutes use common libraries. The formats and frameworks are accessible to other possible users, through a policy of open-source software developments [6, 21]. Data privacy and security are enabled by proprietary front-end layers. Further compatibility with ITER and the IMAS data storage architecture is pursued with EUROfusion [22], including the previously mentioned merging with synthetic IR image data. The compatibility to ITER formulated needs in terms of machine protection based on IR viewing system is also tentatively addressed, as reported in [23]. All this activity in coordination with W7-X, EUROfusion and ITER teams corresponds to the initiation of a cross-fusion institute wall protection community.

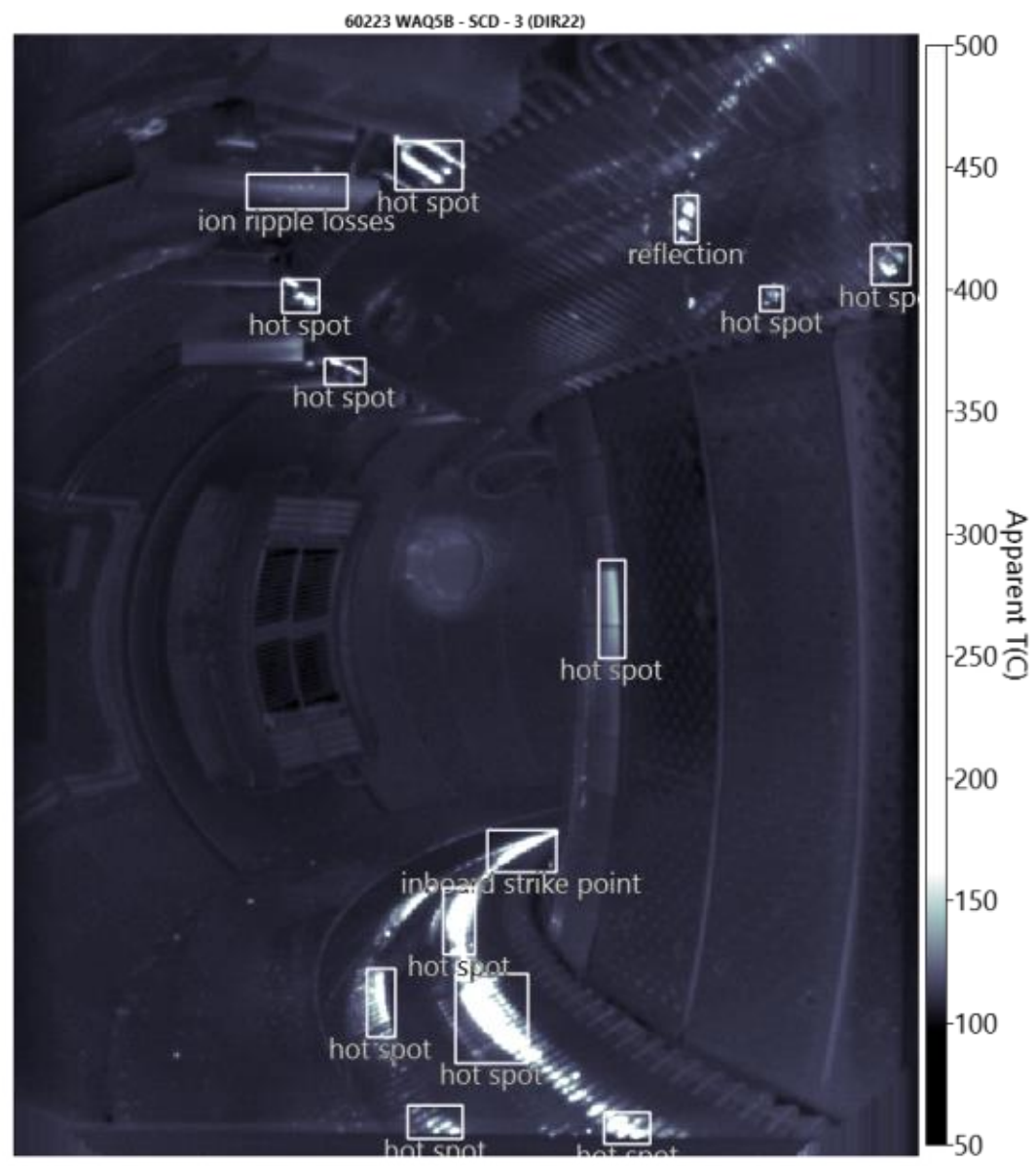

Fig. 5: Bounding boxes surrounding thermal events being detected during WEST discharge #60223.

Various detection algorithms have been employed over time, namely mask-RCNN, then faster RCNN [24]. The YOLO family of detectors is considered for future developments, since it can provide a good trade-off between detection performance and inference speed. Bounding boxes of detected events are given in Fig. 5 for discharge #60223. The “hot spot” label is the catch-all class which fetches thermal events that could not be classified in a known category. These detectors operate in RT, and offer a smart alternative to rigid ROIs. After the discharge, the thermal events are stored in the WEST database. They are ranked according to a severity index, which is basically the ratio of peak temperature to allowable temperature. The ranked thermal events are accessible through a web-based interface, and also by the THERMAVIP software suite [5, 6].

The next tool is the strikeline characterisation tool. It aims at providing the strikeline curvature and angle within the image, for determining whether a strikeline is considered normal or pathologic. Curvature and angle are simple indicators that would be used unconsciously by the PPOs to decide whether a given strikeline is normal or not. Normal strikeline are merely processed for temperature and flux, while pathologic strikelines are further analysed manually for deeper understanding. The curvature is rated on an integer scale (from 0 to 3), and the angle is given in degree, relative to the image main axis. The strikeline tool operates downstream the thermal event detector, on the bounding boxes having been labelled as “strikeline”. It is a constrained U-net [25, 26], which both produces the strikeline skeleton, as illustrated by the connected boxes in Fig. 6 above, and also the two quantities of the angle and curvature. Both numbers are then compared to ranged standards, and the strikelines classified according to whether they are normal or not. The term "constrained" refers to the fact that the bottleneck of the U-Net, in addition to feeding the decoder for pixel-wise segmentation of the hottest points, is also passed through a dense network that regresses the strikeline's angle and curvature, so that these two branches are jointly optimized through a single composite loss and the encoder is forced to preserve physically meaningful information during down sampling.

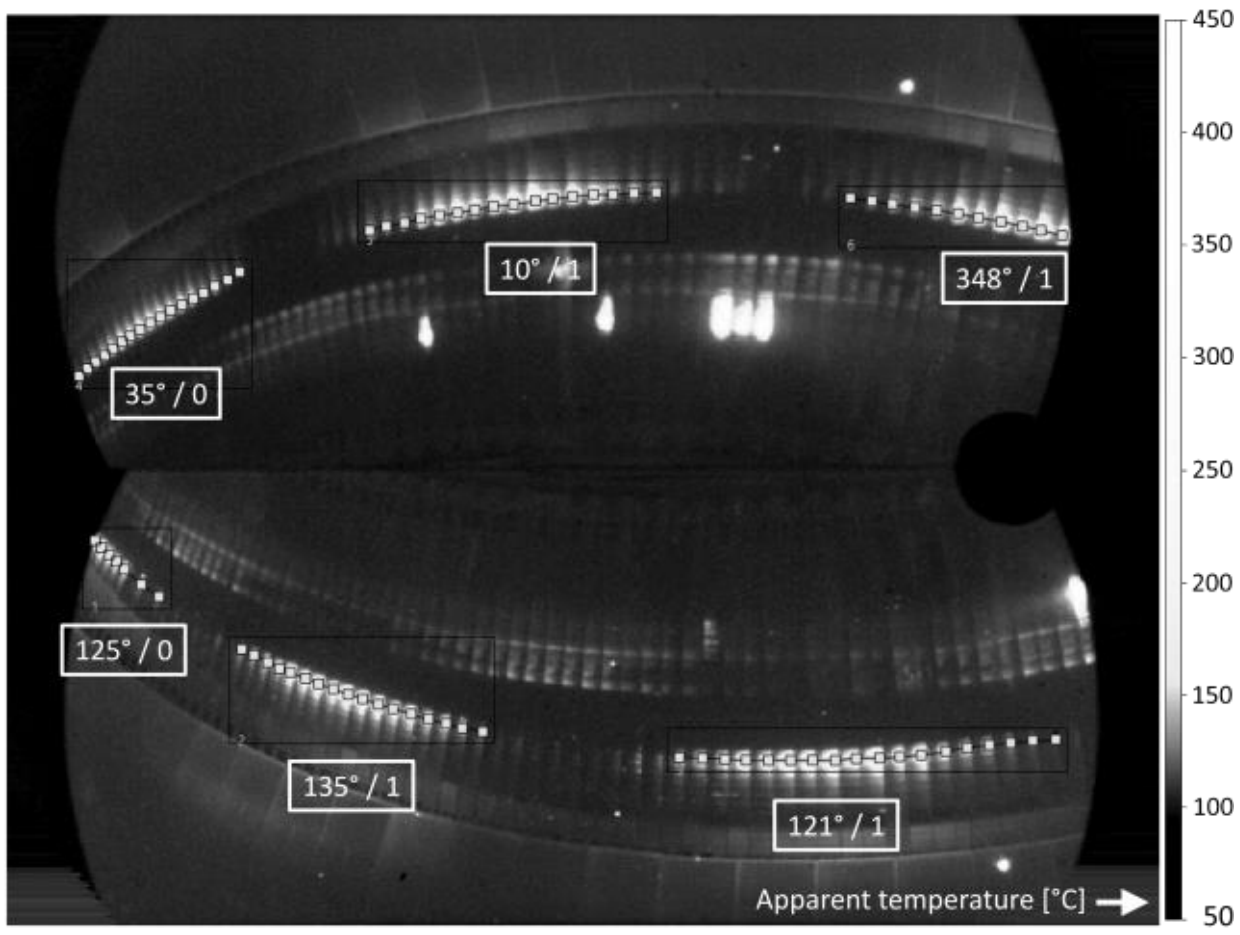

Fig. 6: Extraction of strikelines properties during #57396: segmentation of peak thermal regions, angle and curvature (Angle in ° / Curvature index from 0 to 3).

Another RT AI-based detector is the LH arc detector. It detects electrical arcs in the FoV looking at the waveguides of the LH launcher. The IR images provide the most relevant signature of electrical arcs, better than the copper impurity signal or the antenna reflection coefficient. The arc detector is a lightweight neural network acting on the image frame. It flags the images where an arc is detected, without even producing the bounding box (Fig. 7). The dataset comprises 127 films acquired on the LH1 line of sight, of which 57 contain at least one electric arc event, against a strongly imbalanced background of non-arc frames (0.64% positive frames overall). This imbalance is addressed through a Gaussian temporal sampling scheme at frame-generation time, which biases the retention of negative frames according to their temporal proximity to labelled arc events, combined with a weighted random sampler and a positive class weight in the training loss; no image-level augmentation is applied. The detection threshold is selected to maximize the $F_\beta$ score with $\beta = 0.5$, which weights precision more heavily than recall. This choice reflects the fact that the detector's output feeds directly into the PCS to trigger an automatic reduction of LH power: a false positive therefore has a direct operational cost, in the form of an unnecessary power cut and disruption of the plasma discharge, motivating a threshold that limits spurious detections while the class-weighted sampling and loss prevent the rare arc class from being ignored.

On a held-out validation set (5,958 frames from 25 movies from pulses disjoint from the training set ones, 11 distinct arc events in total), the detector achieves an accuracy of 99.9% and an F1-score of 0.941 for the arc class, with a frame-level false-negative rate of 3.0% (1 missed frame out of 33 positive frames). At the event level, the operationally relevant quantity for a safety-oriented detector, all 11 arc events in the validation set were detected by at least one flagged frame, corresponding to an event-level miss rate of 0% on this validation set. To our knowledge, no quantitative estimate of the characteristic timescale of arc-induced damage to the antenna grill is currently available; despite its 20 ms detection latency, set by the 50 Hz acquisition rate, the image-based detector is regarded by LH experts as the most reliable arc detection tool currently at their disposal, in the absence of a faster or more accurate alternative.

The overall data processing framework being established, building further specialized AI-based image detectors becomes straightforward. The UFO detector (called UFOund) is developed as a result of the need to count UFOs, following the 2022 high fluence campaign [27, 28]. UFOund is made of three convolutional blocks followed by an average pooling layer and a linear feed-forward model. The total number of trainable parameters of UFOund is 151442. This small size enables it to run on small GPUs in near RT during operation. The dataset is generated from 60 pulses of the C9 campaign, leading to the manual annotation of 295 infrared movies. The total number of annotated frames is 134712, among which 7603 contained UFOs (5.64% of the total number). The model has been implemented in Python using PyTorch, and trained and evaluated on a dedicated server equipped with three NVIDIA A30 GPUs (each with 24Gb of VRAM).

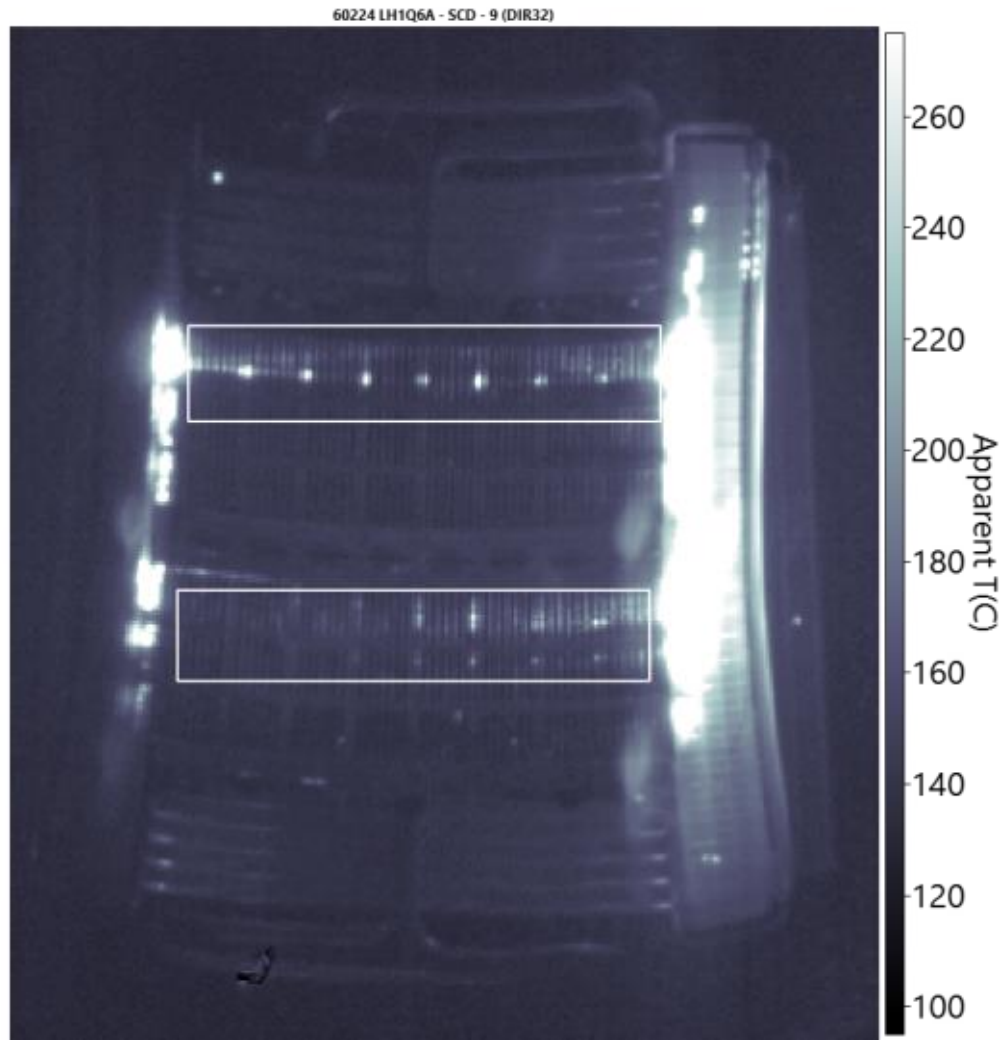

Fig. 7: Electrical arcs detected by a compact arc detector on the LH1 lower-hybrid heating antenna. The bounding boxes surrounding the arcs have been drawn manually to guide the reader's eye, as the detector does not calculate the boxes.

Following the same principle, a preliminary RE detector is currently under development, illustrating how quickly new event-specific detectors can be built on top of the existing infrared pipeline: an initial prototype was developed in approximately two days. The RE detector targets the synchrotron light emission of the RE beam, whose spectrum includes some emission in the waveband of the IR detector (3.9 µm), making it visible on the tangential field of view. This detector is at an early stage of development; dataset construction, training, and performance characterization are left for future work.

### 2.3. The post discharge thermal event diagnostic tool

Further ability to integrate the elaborate knowledge of the WEST thermal scenes is researched, in relation with the plasma session general parameters. This knowledge is virtually defined as the collective memory of the PPOs. It is too complex to be formulated as a list of rules. There are too many dependencies, qualitative judgements, and uncertainties. Large Language Models (LLMs) emerge as a possible candidate technique to integrate such complex knowledge.

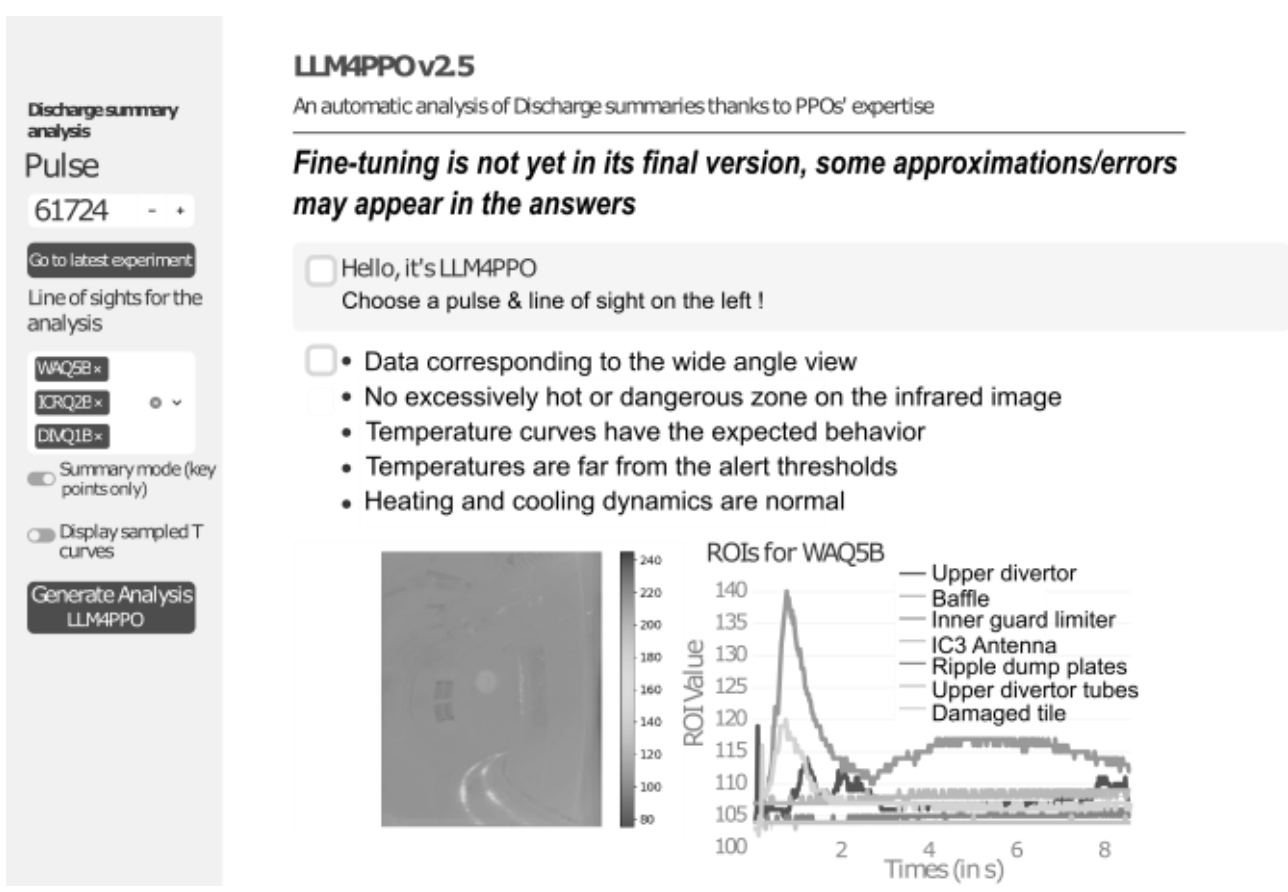

Fig 8: LLM4PPO result for #61724 (The original screen output has been edited with larger fonts, and the prompt and answer have been translated from French to English).

Practically at WEST, LLM developments are based on the discharge summary. The summary is the IR viewing system dashboard. It is the first tool used by the PPOs for analysing the thermal scenes of a plasma experiment once finished. The PPO dashboard is a one-screen summary view of all FoV of a given discharge. There are two sub-tables for each FoV: the “maxed” view of the FoV, which contains an image where each pixel is actually the maximum intensity of each pixel over the time of the discharge, and the time traces of the ROIs (temperature curves as a function of time). The “maxed” image is an artificial image constructed from the whole temperature movie after the discharge, allowing to give a fast and synthetic representation of the most relevant hot spots having occurred in a FoV during the discharge. Two additional views are provided at the bottom of the dashboard, namely the discharge key parameters (plasma current, additional power for each heating system, plasma density) allowing to relate possible hot spots to notable events of the discharge.

An LLM training database of 155 entries is created from these dashboards, associated to textual comments from the PPOs. The multimodal LLM Llava 1.6 [29] is used, which is an appropriate compromise in terms of performance and confidentiality, while still being an open-source model. The interfacing of the dashboard data to the LLM uses a dedicated transfer tool, since the LLM only works with series of “tokens in input” (lexical or data units). An original projection module for tokenisation interface is trained, allowing to associate both IR images, temperature traces and textual content in a suite of tokens, while staying within the limit of the maximum number of tokens (4096). The model has 13 billion parameters. It is fine-tuned on CEA/CNRS national supercomputers. Once fine-tuned, the model is repatriated to WEST, and takes the name LLM4PPO [30]. The inference is performed on WEST data servers a few minutes after the discharge. A typical result of LLM4PPO is provided in Fig. 8. The pulse number and FoV are selected on the left-hand side table, and the analysis results is displayed on the right-hand side after a few seconds in a textual formal. The sentences point at either a normal or abnormal thermal scene and sequence, providing further analysis directly understandable by PPOs.

## 3. RESULTS

### 3.1. Performance of machine learning based processes

Like health and safety monitoring, quantifying performance for matters of operation safety and machine protection is a broad field in itself. Poor performance is obvious when mishaps and accidents do occur frequently. By contrast, performance quantification attracts little attention when plasma operation runs well. Some automatized performance processes are also tuneable with numerical parametric settings. These settings give users the choice of catching all events, at the price of risking some false positives. Opposite settings favour avoiding false positives, accepting the risk of missing some detections. Finally, performance quantification is also a deeply specialized scientific domain on its own. Full books on this topic of performance quantification are available in the literature [31]. There are many possible metrics, depending on the usage or client:

- Loss functions are used for the training phase of the model. Loss function metrics are basically numerical distances, taken between two data points.
- Metrics such as precision, recall, and F1 score are used when it comes about comparing the performance of various possible candidate classification processes. There are known shortcomings to these indicators, for example for imbalanced classification processes. More elaborated indicators are possible (precision versus recall curve, receiver operating characteristic analysis). These tools are specialist’s tools and address experts’ communities.
- And finally, user metrics may be defined, in coordination with clients and users. For example, a user metric can be derived from a polling satisfaction test. User metrics are simple and easily explainable, but may be not as rigorous as the previous ones.

The thermal event detector achieves, on a left-out test set and at an intersection over union (IoU) threshold of 0.5, a recall of 68.07%, a precision of 76.72%, and a mean average precision (mAP) of 59.70%. Given the machine protection context, these results provide a solid baseline. Future work will focus on improving recall, notably through the planned transition from Faster R-CNN to a YOLO-based architecture better suited to real-time, high-sensitivity detection.

For the strikeline characterisation tool described in section 2.2, quantitative results obtained on the test set (synthetic images) demonstrate near perfect segmentation metrics (F1-score > 98%) and low regression errors (mean absolute error < 1° for angle and < 0.05 for curvature). Beyond the synthetic-data evaluation, the strikeline descriptor was also validated on real WEST infrared images [26], without any fine-tuning on experimental data. On two real strikelines, most localization errors were found within a one-pixel distance of the expected position, indicating that residual discrepancies are primarily positional rather than complete misses, despite the network never having been exposed to real infrared noise or texture during training. This domain shift was attributed to differences between the synthetic and experimental image statistics, and incorporating annotated real images into the training set was identified as a direction for improving robustness under operating conditions. The tool was then used satisfactorily by the users, although a user defined metric is yet to be developed.

UFOund achieves a balanced accuracy of 0.78 and a F1-score of 0.67 on an unseen test set with a detection threshold of 0.95. UFOund's post-pulse use by the PFC Protection Officers has been positively received; this deployment should be regarded as an initial operational proof of concept and a decision-support tool that accelerates identification of candidate UFO events, rather than one replacing PPO judgment. Performance is expected to improve as the annotated database grows. Class imbalance (5.64% positive frames) is handled at two stages [28]: negative frames are subsampled at dataset-construction time with a Gaussian temporal weighting favouring frames close to positive ones, and positive/negative samples are reweighted by inverse class frequency during training. The train/validation/test splits in [28] are built at the movie level, with each split containing at least one movie per pulse to preserve view diversity; movies from the same pulse can therefore appear in different splits. This is not strictly pulse-disjoint and may lead to an optimistic bias in the reported metrics, which is identified as a direction for future validation work.

The quantification of the LLM4PPO performance is even more complex to assess. There is no widely accepted “ground truth” to compare with. The quantification of the expert's system performance relies on evaluating four successive versions of LLM4PPO (v1.0, v1.5, v2.0, v2.5), combined with independent PPO human notations collected through a positive/neutral/negative feedback scale [30, 32]. Version 1.0 is the initial implementation, trained on approximately 70 annotated examples; v1.5 added further fine-tuning data; v2.0 changed the representation of temperature curves in the prompt from image to text; v2.5 introduced a response-summarization stage. As detailed in [32], the cumulative ratio of non-negative feedback rose steadily over the campaign, exceeding 0.7 in later pulses, with a marked drop in negative responses following v1.5 and a further improvement in non-negative feedback following v2.5, indicating consistent gains across successive versions. The latest version (v2.5) currently exhibits a 30% rate of incorrect responses (negative feedback rate given by the PPOs during their test use of LLM4PPO). Given the known tendency of LLMs to hallucinate, this residual error rate is not negligible in a machine-protection context, where an erroneous assessment could mislead operational decisions if taken at face value. For this reason, LLM4PPO is used strictly as an assistant to the PPO: it does not act automatically nor feed into any protective or control system, and the PPO retains full responsibility for the final assessment. The tool is still being improved by training, i.e., by augmenting the training database using PPO expert analyses.

### 3.2. System performance

The infrared viewing system is a critical component to the wall monitoring. Both IR and wall monitoring systems availability and reliability are monitored automatically during the experimental campaigns. The indexes are calculated based on set of discharge data, meeting power and relevance criteria (additional power > 10 kW, elimination of technical tests and failed discharges from the discharge set used for the statistics). The availability index $a_v$ is calculated based on the presence of recorded temperature, both for raw data and ROI security data which comes as a duplication. The presence of the ROI security data is linked to the security watchdog data, also in the signal database. The index come initially as a number between 0 and 1, and is converted as a percentage.

$$a_v = \left\{\frac{N_{data}}{N_{set}}\right\}_{set}$$

During the experimental campaign C9 (Jan-Apr 2024), the availability of the wall protection ($a_v$) is of 94.1%, and 98.4% for C10-C11. The availability index reflects a state where the whole chain of wall protection is fully operational, which is an outstanding availability. The partial availability, defined as the state where one or several safety components do not operate normally, is even larger. Partial availability still allows plasma operation, while certain wall safety functions might not be active. 15.6 % of the discharges entered the active control region, meaning that automatic wall protection measures were activated.

For the same set of discharges, the reliability index counts the good discharges having been lost as a consequence of a malfunction of the wall monitoring system.

$$r_e = 1 - \left\{\frac{N_{lost}}{N_{set}}\right\}_{set},$$

where $N_{lost}$ is the number of well-running discharges having been lost as a consequence of a malfunction of the wall monitoring system (false positive or technical glitch causing undue discharge termination). Only 1 discharge out of 1389 is lost as a result of the hard-wired wall protection system, because the automatic feedback control failed. This shows a >99.9% reliability of the wall protection system for the considered time interval. This outstanding reliability shall still be confirmed during future experimental campaigns.

## 4. CONCLUSION

The WEST Wall monitoring system is developed by incorporating a-priori knowledge either from scientific models or phenomenological know-how recorded as annotated databases. This allows making advanced usage of dense information being hardly accessible in the extensive 2D+time infrared recordings of in-vessel IR viewing systems. The IR data is supplemented with signals coming from other diagnostics through data merging, giving access to heat flux quantities which contribute to the safe operation of actively cooled components. The relevant thermal event and heat flux indicators are used both in real time through machine actuators, and also post discharge for expert's analysis toward adjusting the discharge plan during experimental sessions. No critical wall power event happened during C9 to C11 (2024-2025, about 13 hours of plasma cumulated). While it cannot be demonstrated that active wall protection enabled smooth operation, and especially the new plasma duration record of 1337s, the active wall protection as a whole plausibly contributed to obtaining these records by preventing wall hot spots to become critical during the campaigns.

## ACKNOWLEDGEMENTS

Parts of this work have been carried out within the framework of the EUROfusion Consortium, funded by the European Union via the Euratom Research and Training Programme (Grant Agreement No 101052200 — EUROfusion). Views and opinions expressed are however those of the authors only and do not necessarily reflect those of the European Union or the European Commission. Neither the European Union nor the European Commission can be held responsible for them.